\documentclass[conference]{IEEEtran}
\IEEEoverridecommandlockouts

\usepackage{cite}
\usepackage{amsmath,amssymb,amsfonts}
\usepackage{algorithmic}
\usepackage{graphicx}
\usepackage{textcomp}
\usepackage{xcolor}
\usepackage{booktabs}
\usepackage{multirow}
\def\BibTeX{{\rm B\kern-.05em{\sc i\kern-.025em b}\kern-.08em
    T\kern-.1667em\lower.7ex\hbox{E}\kern-.125emX}}
\renewcommand{\figurename}{Figure} 
\begin{document}

\title{Scaffolding Self-Transduction with Alignment Attention Mask for Aligner-Encoder ASR}

\author{
  \IEEEauthorblockN{Jaeyoung Lee, Masato Mimura}
  \IEEEauthorblockA{
      \\
    \textit{NTT, Inc., Japan}
  }
}

\maketitle

\begin{abstract}
End-to-end ASR systems must jointly solve speech-text alignment and token prediction from sequence-level supervision.
We study this problem in Aligner-Encoder ASR, where alignment is performed inside encoder self-attention and can thus be directly intervened on with an attention mask.
We use forced alignment to build an attention mask that confines selected encoder self-attention heads to a narrow window around the alignment path. The mask is faded out over training, leaving a deployable model with an unmodified inference architecture; alternatively, kept active at training and inference, it serves as an oracle that measures recognition performance once alignment is externally resolved. On LibriSpeech, the faded alignment mask improves test-clean/test-other WER from 3.9/6.4 to 3.5/5.9 over an Aligner baseline. The oracle masks are far stronger, reaching 1.2/2.9 WER, which can be thought of as an upper-bound accuracy in the 100M-class Conformer regime without an external language model. This indicates that alignment inference is a major bottleneck in Aligner-Encoder ASR, and suggests that improved alignment mechanisms may offer substantial headroom for end-to-end ASR.
\end{abstract}


\section{Introduction}
\label{sec:intro}
End-to-end ASR systems differ in how they resolve the latent alignment between
acoustic frames and output tokens. CTC~\cite{graves2006ctc} and
RNN-T~\cite{graves2012rnnt,transformertransducer} marginalize monotonic paths, while attention-based encoder-decoder (AED) models
estimate label-synchronous context through decoder
attention~\cite{chan2016las,watanabe2017hybrid}. Recent large-scale
ASR models also use this AED formulation
\cite{whisper,owsm}. In all cases, training requires jointly solving
alignment inference and token prediction from sequence-level supervision,
mapping an acoustic sequence of length T to a shorter token sequence of length U.

The Aligner-Encoder~\cite{stooke2024aligner} is a recently proposed model that performs this alignment within the encoder.
It dispenses with both the transducer lattice and decoder cross-attention, relying on the encoder self-attention to reorganize frame-level representations into token-level ones, so that the $u$-th token is predicted from encoder output position $u$.
Since the frame-to-token transduction is carried out within the encoder's own self-attention, we call this \emph{self-transduction}.
Alignment is thus performed within the encoder, and recognition and alignment are learned jointly in a single self-attention stack, yielding fast and lattice-free decoding.
The trade-off is that alignment is not guaranteed by the training criterion: the self-attention itself must learn it, using sequence-level supervision alone.
Prior work reports that this alignment emerges abruptly in a few upper encoder layers~\cite{stooke2024aligner}, which can make optimization fragile and can cause accumulated deletion errors on long utterances.

In CTC and RNN-T, alignment is handled indirectly by summing over possible monotonic paths in the training loss.
In Aligner-Encoder, frame-to-token alignment is performed by encoder self-attention itself, making the alignment field directly maskable.
We exploit this property with an \emph{alignment attention mask} to scaffold self-transduction: an external forced-alignment path is converted into an attention mask applied to selected encoder self-attention heads.
For those heads, each token-position query is allowed to attend only near the corresponding frame region on the path.
To our knowledge, this is the first work to transfer an external alignment path directly into the encoder self-attention that performs the alignment, and to use oracle masks to measure the alignment-resolved performance ceiling of that architecture.

We evaluate the alignment attention mask in two regimes: a \emph{faded} regime, where the mask is gradually removed during training, and an \emph{oracle} regime, where the mask remains active during both training and inference to measure recognition when alignment information is externally supplied.
In our experiments, we observe that the faded band can improve recognition while requiring no external aligner at inference, while oracle masks reveal much larger gains when alignment is supplied.
These results indicate that alignment remains a substantial bottleneck in Aligner-Encoder ASR and a promising direction for improving end-to-end ASR in general.

\begin{figure*}[ht]
\centering
\includegraphics[width=\textwidth]{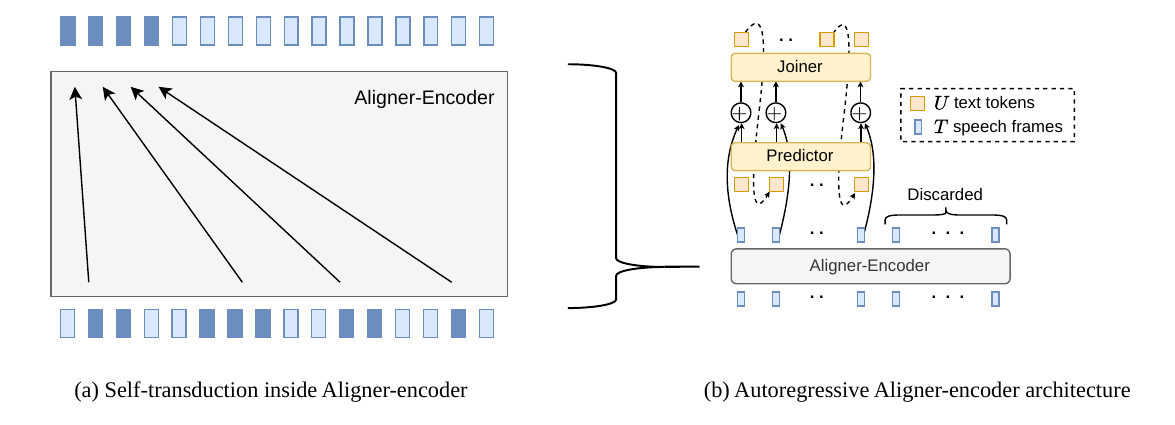}
\caption{Autoregressive Aligner-Encoder architecture. The encoder produces $T$ frame-indexed representations, while the reference transcript contains $U$ target-token positions. The first $U$ encoder positions are paired one-to-one with the target tokens, and each token $y_u$ is predicted by a joiner that combines encoder state $h_u$ with an autoregressive predictor state computed from $y_{<u}$. The remaining encoder positions $h_{U+1:T}$ are discarded by the final Aligner prediction path. This one-to-one pairing requires encoder self-attention to route each token's acoustic evidence into its corresponding token-indexed position.}
\label{fig:aligner}
\vspace{-0.5em}
\end{figure*}

\section{Related Work}
\label{sec:related}

\subsection{Aligner-Encoder ASR and internal self-transduction}
Aligner-Encoder ASR predicts each token $y_u$ from the corresponding encoder
position $u$, requiring encoder self-attention to reorganize frame-level acoustic
representations into token-indexed ones~\cite{stooke2024aligner}. We refer to
this encoder-internal frame-to-token reorganization as \emph{self-transduction}.
Prior work reports that this alignment emerges abruptly in a small
number of upper encoder layers, rather than being uniformly distributed
throughout the network. Our method targets this alignment-formation
stage by constraining selected self-attention heads with externally estimated
alignment information.

This encoder-side formulation makes Aligner-Encoder a natural target for
attention-mask intervention employed in this study: if alignment is performed by encoder self-attention,
then constraining selected heads directly constrains the mechanism that performs
self-transduction.

\subsection{Alignment guidance and alignment knowledge transfer}
A related line of work uses alignment or timing information to make end-to-end
ASR training and decoding more structured. Monotonic attention~\cite{raffel2017monotonic}
and MoChA~\cite{chiu2018mocha} impose monotonic constraints on decoder-side
attention for online or linear-time decoding, while triggered
attention~\cite{moritz2019triggered} uses CTC-derived trigger points to activate
an attention decoder at selected frames.
Supervised-attention methods use external alignments to shape decoder attention~\cite{yang2022supervised}, alignment-knowledge distillation guides online attention emission behavior~\cite{inaguma2023akd}, and self-knowledge distillation can shape CTC spike timing~\cite{kim2024skd}.
Alignment
information has also been used to restrict or reduce transducer training, for
example by limiting the RNN-T lattice with external forced alignment or CTC
guidance~\cite{arrnnt,ctcguided}.

These methods show that alignment information can improve end-to-end ASR beyond conventional sequence-level supervision. However, they generally guide decoder attention, emission timing, CTC criteria, or transducer lattices. In contrast, we inject an externally estimated alignment path as an attention mask into the encoder self-attention that performs Aligner-Encoder self-transduction, as a structural constraint on selected heads rather than as teacher posteriors, hypotheses, decoder-side attention supervision, or a decoding constraint.

\begin{figure*}[t]
\centering
\includegraphics[trim=16 12 32 4, clip, width=\textwidth]{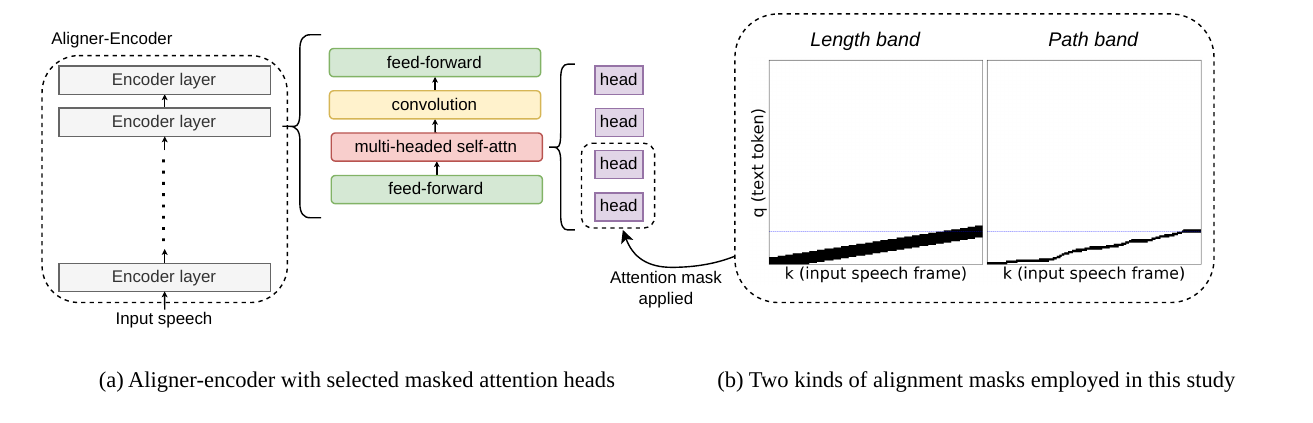}
    \caption{Alignment attention mask for Aligner-Encoder ASR. A monotonic frame--token alignment path, estimated for the reference transcript by an external forced aligner (a pretrained RNN-T in this study), is converted into an attention mask that confines selected encoder self-attention heads to a narrow window around the path (the \emph{path band}); the coarser \emph{length band} instead uses only the endpoint $(T,U)$ to draw a straight band that reveals the target length $U$ and the global frame-to-token compression ratio. These masks are applied to a subset of self-attention heads in a single encoder layer.}
\label{fig:main}
\vspace{-0.5em}
\end{figure*}

\section{The Aligner-Encoder Model}
\label{sec:aligner}

Figure~\ref{fig:aligner} illustrates the autoregressive Aligner-Encoder architecture used in this work. Let $\mathbf{x}$ denote the input speech, an acoustic feature sequence of length $T$. The encoder $E$ maps it to a sequence of encoder representations
\begin{equation}
\mathbf{h}_{1:T} = E(\mathbf{x}).
\end{equation}
For a reference transcript $\mathbf{y}=(y_1,\ldots,y_U)$, including the end-of-sentence token, the model pairs the first $U$ encoder positions with the $U$ target-token positions, as shown in Figure~\ref{fig:aligner}(a). The remaining encoder positions are not used by the final Aligner prediction loss.

The model retains the predictor--joiner interface used in transducer models, but removes the transducer lattice. The predictor $P$ is an autoregressive text network that summarizes the previous output tokens. In our implementation, the predictor is a single-layer LSTM operating over token embeddings. At token position $u$, it produces
\begin{equation}
\mathbf{g}_u = P(y_{<u}),
\end{equation}
where $y_{<u}$ is the token history.

The joiner $J$ combines the encoder representation at the same position with the predictor state. Following standard RNN-T practice~\cite{stooke2024aligner}, we implement the joiner as a single-layer feed-forward network. Specifically, the encoder and predictor representations are first projected to a common dimension and combined via elementwise addition:
\begin{equation}
\tilde{\mathbf{h}}_u = \mathbf{W}_h \mathbf{h}_u + \mathbf{b}_h, \quad
\tilde{\mathbf{g}}_u = \mathbf{W}_g \mathbf{g}_u + \mathbf{b}_g,
\end{equation}
\begin{equation}
\mathbf{z}_u = \mathbf{W}_o \, \phi(\tilde{\mathbf{h}}_u + \tilde{\mathbf{g}}_u) + \mathbf{b}_o,
\end{equation}
where $\mathbf{W}_h,\mathbf{W}_g,\mathbf{W}_o$ are learned weight matrices, $\mathbf{b}_h,\mathbf{b}_g,\mathbf{b}_o$ are bias vectors, and $\phi(\cdot)$ is a nonlinearity (e.g., $\tanh$ or ReLU). The token posterior is then
\begin{equation}
p(y_u \mid \mathbf{x}, y_{<u}) = \mathrm{softmax}(\mathbf{z}_u)_{y_u}.
\end{equation}

Training uses teacher forcing for the predictor history and minimizes the
cross-entropy (CE) loss
\begin{equation}
\mathcal{L}_{\mathrm{AE}} = -\sum_{u=1}^{U} \log p(y_u \mid \mathbf{x}, y_{<u}).
\end{equation}

This differs from RNN-T in that the joiner is not evaluated over a two-dimensional $(t,u)$ lattice and no forward--backward marginalization over alignments is performed. Instead, the model is trained on the one-to-one pairing between encoder position $u$ and output token $y_u$. Consequently, the encoder must learn to route the acoustic evidence for each output token into the corresponding token-indexed position before the joiner is applied. This encoder-internal frame-to-token reorganization is the self-transduction operation targeted by our alignment attention mask.

At inference time, given a partial hypothesis $\hat{\mathbf{y}}_{<u}$, the predictor produces $P(\hat{\mathbf{y}}_{<u})$, and the joiner scores the next token from encoder position $u$:
\begin{equation}
p_\tau(v \mid \mathbf{x}, \hat{\mathbf{y}}_{<u}, u) =
\mathrm{softmax}\!\left(
\frac{1}{\tau} J(\mathbf{h}_u, P(\hat{\mathbf{y}}_{<u}))
\right)_v,
\end{equation}
where $v$ ranges over the output vocabulary and $\tau$ is a decoding temperature. Beam search extends each hypothesis by one token at each encoder position until an end-of-sentence token is emitted.

We rank completed hypotheses with a length-penalized score
\begin{equation}
s(\hat{\mathbf{y}}) =
\sum_{u=1}^{|\hat{\mathbf{y}}|}
\log p_\tau(\hat{y}_u \mid \mathbf{x}, \hat{\mathbf{y}}_{<u}, u)
- \beta \, n(\hat{\mathbf{y}}),
\end{equation}
where $n(\hat{\mathbf{y}})$ is the number of emitted non-EOS tokens and $\beta$ is the length penalty.

\section{Alignment Attention Mask}
\label{sec:mask}

The Aligner-Encoder training objective pairs output token $y_u$ with encoder
position $u$, so speech recognition succeeds only if the encoder has already routed the
corresponding acoustic evidence into that token-indexed position. Prior work
observes~\cite{stooke2024aligner} that this frame-to-token alignment emerges
abruptly in a small number of upper encoder layers, rather than being uniformly
represented throughout the network. We hypothesize that this alignment formation
can be scaffolded by guiding the self-attention heads that participate in this
routing process.
We implement this guidance with an \emph{alignment attention mask}, which forces
selected heads to attend within a narrow band around a frame-to-token alignment
path.
The band is defined formally below.

At the same time, such a mask is a strong constraint because it directly restricts where a query position can gather acoustic evidence.
If imposed on all heads or many layers, it could prevent the encoder from using off-band acoustic, phonetic, or contextual information.
We therefore apply the mask only to selected attention heads in a single encoder layer.
The masked heads provide an alignment-guided route for acoustic evidence, while the remaining heads and layers remain unconstrained and can learn complementary representations.
Figure~\ref{fig:main} provides an overview of the alignment-mask construction and where it is applied in the Aligner-Encoder.

Let $s$ denote the encoder layer to which the mask is applied, and let $\mathcal{H}_{\mathrm{band}}$ denote the selected band heads, with $K=|\mathcal{H}_{\mathrm{band}}|$.
The mask is applied only to these heads.
Unselected heads, other layers, and the remaining model components are left unchanged.

\subsection{Alignment Bands and Masked Attention}

For a center function $c(t)$ that maps each frame $t$ to a token index and a
half-width $b$, we define the band for token query $u$ as
\begin{equation}
\mathcal{B}_u(c,b) = \{\, t\in\{1,\ldots,T\} : |u-c(t)| \le b \,\}.
\label{eq:band}
\end{equation}
Thus, $\mathcal{B}_u(c,b)$ is the set of acoustic frames that token query
$u$ is allowed to attend to.

The \emph{length-conditioned linear band}, or \emph{length band} in short,
uses a linear center,
\begin{equation}
c^{\mathrm{len}}(t) = 1 + \frac{U-1}{T-1}(t-1).
\label{eq:lencenter}
\end{equation}
The resulting band
$\mathcal{B}^{\mathrm{len}}_u=\mathcal{B}_u(c^{\mathrm{len}},b)$
provides the global frame-to-token compression ratio, but no local token timing.
We call this the length band because the only transcript-side information it
uses is the target length $U$; it does not use any token-level timing or
forced-alignment path.

The \emph{full alignment-path band}, or \emph{path band} in short, uses an
externally estimated alignment path. Given the
reference transcript, a forced aligner provides a cumulative token index
$\hat{U}(t)$ for each frame after resampling to the valid encoder-frame
sequence. We define
\begin{equation}
c^{\mathrm{path}}(t)=\hat{U}(t),
\label{eq:pathcenter}
\end{equation}
and
$\mathcal{B}^{\mathrm{path}}_u=\mathcal{B}_u(c^{\mathrm{path}},b)$.
Unlike the length band, the path band follows the local frame-token alignment
path and therefore provides token-dependent timing information. Constructing
$\hat{U}(t)$ requires an accurate external frame-to-token alignment, which we
obtain from a pretrained RNN-T model in this study. Teacher-forcing its predictor on the
reference transcript and running the standard forward--backward pass over the
RNN-T alignment lattice yields, for each encoder frame $t$, the
posterior-expected token index $\hat{U}(t)$, which we take as the path center
$c^{\mathrm{path}}(t)$.

\subsection{Mask Scheduling and Oracle Evaluation}

The mask acts multiplicatively on the pre-softmax attention weights of the
selected heads. For a band $\mathcal{B}^{r}_u$ with
$r\in\{\mathrm{len},\mathrm{path}\}$, it scales the attention score of key $t$
for query $u$ by
\begin{equation}
m^{r}_{u,t}(n) =
\begin{cases}
1, & t\in\mathcal{B}^{r}_u,\\
\max\!\big(\varepsilon, g(n)\big), & t\notin\mathcal{B}^{r}_u.
\end{cases}
\label{eq:maskfactor}
\end{equation}
The scalar $g(n)\in[0,1]$, where $n$ is the training step, is an out-of-band
gate: out-of-band keys retain a fraction $g(n)$ of their attention weights, while
in-band keys are unbiased. For numerical stability, the out-of-band weight is
floored at $\varepsilon=10^{-4}$, so a valid key never receives exactly zero
softmax mass. Thus $g=0$ gives a strongly band-limited bias, $0<g<1$ gives a
soft bias, and $g=1$ removes the mask.

We use each band in two regimes. In the \emph{faded} regime, the mask is gradually faded out during training, and is not used at inference.
We set
\begin{equation}
g(n)=
\begin{cases}
0, & n < n_1,\\[2pt]
\dfrac{n-n_1}{n_2-n_1}, & n_1 \le n \le n_2,\\[6pt]
1, & n > n_2 .
\end{cases}
\label{eq:fade}
\end{equation}
where $n$ is the optimizer step. Before $n_1$, out-of-band keys receive only the
small floor weight $\varepsilon$, so selected heads are strongly biased toward
the band. Between $n_1$ and $n_2$, $g$ increases linearly from $0$ to $1$,
gradually restoring out-of-band attention. After $n_2$, $g=1$ and the mask is
removed, so the model trains and decodes as the original Aligner-Encoder.
This regime tests whether alignment side information can be internalized while
preserving the original inference-time architecture.

In the \emph{oracle} regime, the mask remains active through both training and inference.
We disable the fade and set $g(n)=\varepsilon$, giving a strong but non-degenerate out-of-band penalty.
This setting is intentionally non-deployable because the length band uses the reference target length $U$ and the path band uses a reference-derived alignment path.
It should be noted that this is not teacher-forced evaluation.
Teacher forcing would condition the predictor on the reference prefix $y_{<u}$, making the token history oracle as well.
In our oracle regime, decoding remains autoregressive, and only the target length $U$ or the alignment path is reference-derived.
The \emph{oracle} settings are diagnostic, measuring recognition performance with either the correct global length and slope supplied by the oracle length band or the full frame-token alignment path.

\section{Experiments}
\label{sec:exp}

\subsection{Experimental Setup}
We evaluate on LibriSpeech 960\,h~\cite{librispeech} and a five-language Common
Voice 16.1 setting~\cite{commonvoice} covering English, German, French, Spanish,
and Italian with $4{,}133$\,h of training audio. LibriSpeech uses speed
perturbation, both settings use SpecAugment~\cite{specaugment}, and performance
is reported as WER on test-clean/test-other.

All systems use the same Conformer-L encoder~\cite{conformer} with approximately
$115$M parameters, $17$ layers, and $8$ attention heads. The Aligner predictor is
a one-layer LSTM with $512$ hidden units, and the joiner has $640$ hidden units.
We use BPE vocabularies of size $1024$ for LibriSpeech and $2000$ for the
multilingual Common Voice system. All Aligner models include an auxiliary
InterCTC loss~\cite{interctc} of weight $0.1$ at encoder layer $12$. Path-band
alignments are obtained from an RNN-T teacher trained on the same corpus as the
corresponding Aligner-Encoder student.

Models are trained for $80$ epochs on LibriSpeech and $35$ epochs on Common
Voice, with $20$k warmup steps, peak learning rate $0.0020$, and an effective
batch size of approximately two hours of audio. Final models are obtained by
averaging the $10$ best LibriSpeech checkpoints and the $5$ best Common Voice
checkpoints. Unless otherwise stated, decoding uses beam width $4$, temperature
$\tau=1.25$, and length penalty $\beta=0.5$.

For alignment masks, the path band uses half-width $b=1$, while the length band
uses $b=5$. Faded masks use $[n_1,n_2]=[2000,10000]$, and oracle masks keep
$g(n)=10^{-4}$. The masked layer and number of band heads are found optimal at
$(s,K)=(17,1)$ for faded masks and $(s,K)=(15,4)$ for oracle masks.

\subsection{Main Results}
Tables~\ref{tab:libri} and~\ref{tab:cv} report the main results. On LibriSpeech,
as shown in Table~\ref{tab:libri},
the faded length band yields similar performance to the baseline, with $3.9/6.4$
WER on test-clean/test-other, indicating that a transient length-only scaffold
is not sufficient to improve the deployable model. In contrast, the faded path
band improves WER to $3.5/5.9$, showing that local alignment-path information can
be partially internalized during training. The oracle bands reveal substantially
greater potential for improvement: the oracle length band reaches $2.3/5.3$,
showing that the correct target length $U$ and global frame-to-token slope are
themselves valuable when available at inference, and the oracle path band further
improves WER to $1.2/2.9$.

\begin{table}[t]
\centering
\caption{LibriSpeech results (WER, test-clean/test-other).}
\vspace{-0.5em}
\label{tab:libri}
\footnotesize
\begin{tabular}{l cc}
\toprule
System & clean & other \\
\midrule
\textit{RNN-T teacher}        & 2.2 & 5.1 \\
\midrule
\textit{Aligner baseline}             & 3.9 & 6.4 \\
\textit{+ length band (faded)}  & 3.9 & 6.4 \\
\textit{+ path band (faded)}    & 3.5 & 5.9 \\
\midrule
\textit{+ length band (oracle)} & 2.3 & 5.3 \\
\textit{+ path band (oracle)}   & 1.2 & 2.9 \\
\bottomrule
\end{tabular}
\vspace{-0.5em}
\end{table}

\begin{table}[t]
\centering
\caption{Common Voice per-language WER.}
\vspace{-0.5em}
\label{tab:cv}
\footnotesize
\setlength{\tabcolsep}{3pt}
\begin{tabular}{l cccccc}
\toprule
System & en & de & fr & es & it & Average \\
\midrule
\textit{RNN-T teacher} & 16.0 & 8.1 & 12.9 & 8.2 & 9.3 & 10.9 \\
\midrule
\textit{Aligner baseline}           & 17.9 & 9.5 & 14.3 & 9.3 & 10.9 & 12.4 \\
\textit{+ path band (faded)}  & 17.5 & 9.0 & 13.9 & 9.1 & 10.4 & 12.0 \\
\midrule
\textit{+ path band (oracle)} & 8.5  & 4.6 & 7.2  & 3.7 & 4.3  & 5.7  \\
\bottomrule
\end{tabular}
\vspace{-0.5em}
\end{table}

\begin{table}[t]
\centering
\caption{LibriSpeech WER by utterance-duration bucket (test-clean/test-other).}
\vspace{-0.5em}
\label{tab:duration}
\footnotesize
\setlength{\tabcolsep}{3pt}
\begin{tabular*}{\columnwidth}{@{\extracolsep{\fill}}llrrrr@{}}
\toprule
Set & System & $<11$s & 11--17s & $>17$s & Overall \\
\midrule
\multirow{6}{*}{\textit{clean}} & \textit{RNN-T teacher} & 2.3 & 1.7 & 2.4 & 2.2 \\
\cmidrule(l){2-6}
& \textit{Aligner baseline} & 2.5 & 2.1 & 11.7 & 3.9 \\
& \textit{+ length band (faded)} & 2.5 & 2.0 & 11.8 & 3.9 \\
& \textit{+ path band (faded)} & 2.6 & 1.9 & 9.3 & 3.5 \\
\cmidrule(l){2-6}
& \textit{+ length band (oracle)} & 2.3 & 1.6 & 3.7 & 2.3 \\
& \textit{+ path band (oracle)} & 1.3 & 0.9 & 1.3 & 1.2 \\
\midrule
\multirow{6}{*}{\textit{other}} & \textit{RNN-T teacher} & 5.5 & 4.3 & 4.3 & 5.1 \\
\cmidrule(l){2-6}
& \textit{Aligner baseline} & 5.9 & 4.6 & 12.8 & 6.4 \\
& \textit{+ length band (faded)} & 5.9 & 4.4 & 12.9 & 6.4 \\
& \textit{+ path band (faded)} & 5.7 & 4.2 & 10.8 & 5.9 \\
\cmidrule(l){2-6}
& \textit{+ length band (oracle)} & 5.5 & 4.2 & 6.4 & 5.3 \\
& \textit{+ path band (oracle)} & 3.1 & 2.3 & 2.2 & 2.9 \\
\bottomrule
\end{tabular*}
\vspace{-0.5em}
\end{table}

The RNN-T teacher is stronger than the deployable Aligner systems with faded masks, achieving $2.2/5.1$ WER, but this comes with
computationally heavier transducer training and decoding. The faded path band
instead transfers part of the teacher's alignment knowledge into an
Aligner-Encoder student that retains the Aligner-Encoder's CE-based objective without transducer marginalization and lower-cost
token-indexed decoding: excluding the shared encoder and beam factor, the
decoding path is reduced from RNN-T's $O(T+U)$ frame-label traversal to $O(U)$
joiner evaluations.

As shown in Table~\ref{tab:cv}, The same trend holds in the five-language Common Voice setting.
The faded path band improves all five languages, reducing average WER from $12.4$ to $12.0$,
while the oracle path band gives a much larger gain, reducing average WER to
$5.7$. These results suggest that when the local frame-token correspondence is
supplied, the remaining token-prediction problem becomes substantially easier
across languages. In summary, length information has large oracle value but weak
transfer as a transient scaffold, whereas path information provides both a
deployable improvement and a much stronger alignment-resolved oracle gain.

\begin{figure}[ht]
\centering
\includegraphics[width=0.7\columnwidth]{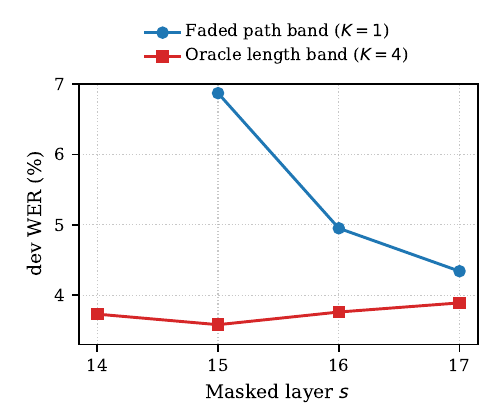}
\vspace{-1.0em}
\caption{Masked-layer ablation on LibriSpeech (overall dev WER). The faded path band ($K{=}1$) improves as the mask moves to deeper layers, whereas the oracle length band ($K{=}4$) is best near layer $15$.}
\label{fig:layer}
\vspace{0.5\baselineskip}
\includegraphics[width=0.7\columnwidth]{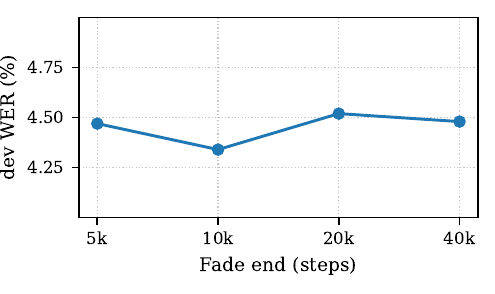}
\vspace{-1.0em}
\caption{Fade-schedule ablation for the faded path band on LibriSpeech. The mask fades from $2$k steps to the listed end step.}
\label{fig:steps}
\vspace{-0.5em}
\end{figure}

\subsection{Utterance-Duration Breakdown}
Table~\ref{tab:duration} breaks down LibriSpeech WER by utterance duration. Two
trends are notable. First, the faded path band improves the baseline Aligner
model most clearly on long utterances. On test-clean, the $>17$s bucket improves
from $11.7\%$ to $9.3\%$, while the short-utterance bucket changes little; on
test-other, the same long-utterance bucket improves from $12.8\%$ to $10.8\%$.
This suggests that the faded path band transfers not only local alignment hints,
but also part of the RNN-T teacher's robustness to long-utterance alignment.
Second, the oracle path band nearly removes the duration dependence: WER becomes
similar across short, medium, and long buckets, especially on test-clean. Once
the frame-token correspondence is supplied, long utterances are no longer
disproportionately difficult.

\begin{figure}[ht]
\centering
\includegraphics[width=\columnwidth]{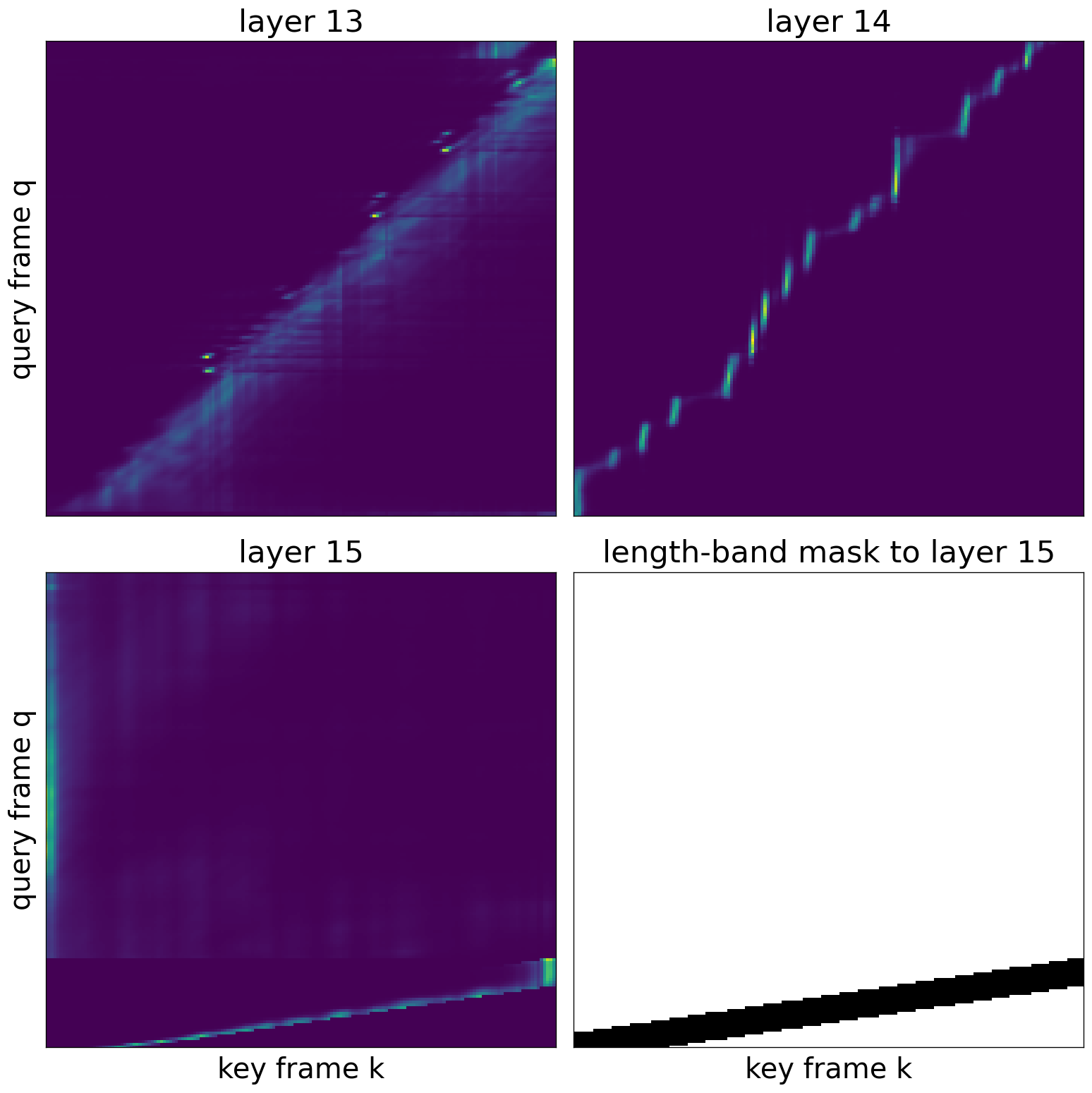}
\caption{Encoder self-attention (query position $q$ vs.\ key position $k$) for one masked head of the oracle length-band model ($s=15$, $K=4$). The first three panels show the head's attention at layers 13--15; the last panel shows the length-band mask applied at layer 15 (black marks the allowed band). A near-diagonal frame-to-token alignment emerges at layer 14.}
\label{fig:attn}
\vspace{-1.0em}
\end{figure}

\subsection{Ablations}
Figures~\ref{fig:layer} and~\ref{fig:steps} summarize the mask-placement and
fade-schedule ablations. The oracle mask performs best around layers
$14$--$15$, consistent with the original Aligner-Encoder
observation~\cite{stooke2024aligner} that frame-token alignment naturally emerges
in a small number of upper encoder layers. Since the oracle mask remains active
at inference, the best placement is therefore the layer where the model already
tends to form its internal alignment.

The faded path band shows a different pattern: with one band head, performance
improves as the masked layer moves upward, with the best result at the final
encoder layer ($s=17$). This is plausible because the mask is eventually removed.
After release, the model must maintain the scaffolded alignment using only the
final recognition objective; placing the mask closer to the final Aligner loss may provide a stronger learning signal for preserving the induced
alignment. Lower layers such as $s\in\{15,16\}$ may be too far from the final loss for this transient cue to be retained as effectively.

The fade-schedule ablation shows that the path band is most effective as a brief
early scaffold. The best setting fades the mask from $2\mathrm{k}$ to
$10\mathrm{k}$ optimizer steps, finishing before the $20\mathrm{k}$-step
learning-rate warmup is complete. Overall, these ablations indicate that
alignment masking works best as a minimal, early intervention: it should provide
enough guidance to seed self-transduction, while leaving the model free to learn
the final unconstrained representation.

\begin{figure}[ht]
\centering
\includegraphics[width=0.8\columnwidth]{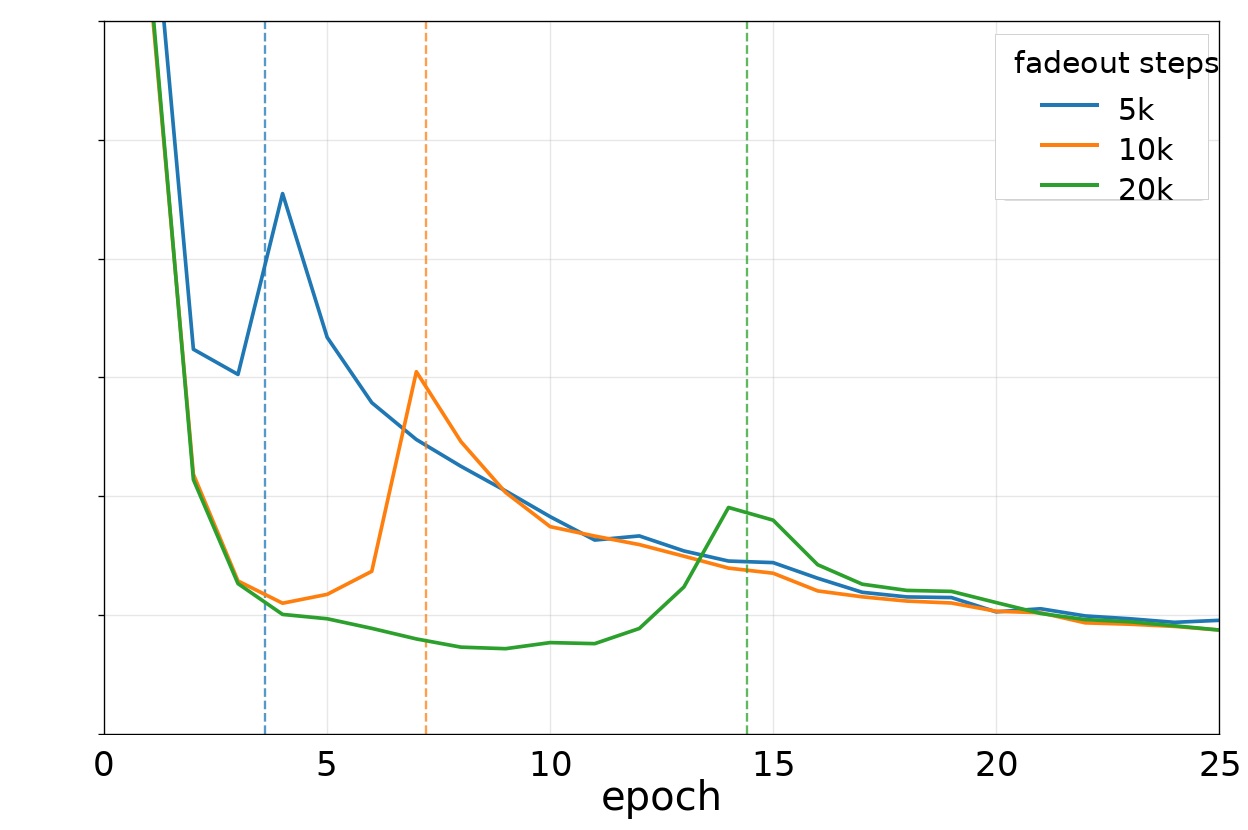}
\caption{Validation-loss trajectories for faded path-band models with different fadeout steps on LibriSpeech. Dashed vertical lines mark the end of each fade window.}
\label{fig:loss-fade}
\vspace{-1.0em}
\end{figure}

\subsection{Emergent Alignment under the Oracle Length Band}
Figure~\ref{fig:attn} visualizes attention patterns under the oracle length-band
condition. When the length band is applied at layer~15, a near-diagonal
alignment pattern emerges in the preceding layer, with boundary-like vertical
transitions roughly matching the target length $U$. This pattern likely arises
because the length band provides only the endpoint $(T,U)$ and global
frame-to-token slope, not local token timing; to make use of the length
constraint, the preceding layer must first reorganize the representation into a
more token-synchronous form.

This may help explain the strong oracle length-band result.
Although the mask does not provide a full path, fixing $U$ and the global slope can encourage a more local self-transduction geometry.
Because the emergent alignment pattern stays near $k\approx q$, each query attends mostly to key positions with similar encoder indices, instead of attending far away along the global frame-to-token slope.
In other words, the model retrieves relevant acoustic evidence through more local self-attention rather than long-range attention, which may benefit long-utterance generalization when relative positional encodings~\cite{shaw2018self} are in use.

\subsection{Effect of Releasing the Alignment Scaffold}
Figure~\ref{fig:loss-fade} shows the validation loss during faded path-band
training. A clear spike appears when the remaining alignment bias is removed at
the end of the fade schedule, when $g(n)$ reaches $1$ in
Eq.~\eqref{eq:fade}. Just before this point, $g(n)$ is already close to $1$, so
out-of-band attention is largely restored and the remaining mask contribution is
numerically small. The spike therefore suggests that this weak residual bias
toward the aligned frame-token region remains functionally important. Thus,
alignment information appears to be a high-leverage training signal: selected
heads continue to use weak alignment guidance near release, while the final WER
improvement indicates that part of this structure is internalized and remains
useful after the mask is removed.

\section{Conclusion}
We presented an alignment attention mask for Aligner-Encoder ASR that transfers
external alignment information into selected encoder self-attention
heads while preserving the model's simple cross-entropy (CE) training and
low-cost inference after fading. Faded path bands can improve recognition
in practice, while oracle masks reveal much larger theoretical gains when the
full alignment path is supplied. These results indicate that alignment inference
remains a substantial bottleneck in Aligner-Encoder ASR, and motivate future work on
non-oracle alignment prediction, alternative alignment sources, and streaming or
long-form extensions.

\newpage

\section*{AI-Generated Content Disclosure}
LLM-based AI tools were used to assist with code editing, LaTeX build
verification, table and figure layout adjustment, and manuscript language
revision. All AI-assisted content, code changes,
citations, numerical values, and rendered outputs were manually reviewed and
verified by the authors, who take full responsibility for the final manuscript.


\begin{thebibliography}{00}
\bibitem{graves2006ctc} A. Graves, S. Fern\'andez, F. Gomez, and J. Schmidhuber, ``Connectionist temporal classification: Labelling unsegmented sequence data with recurrent neural networks,'' in \emph{Proc. ICML}, 2006, pp. 369--376.
\bibitem{graves2012rnnt} A. Graves, ``Sequence transduction with recurrent neural networks,'' 2012, arXiv:1211.3711.
\bibitem{chan2016las} W. Chan, N. Jaitly, Q. V. Le, and O. Vinyals, ``Listen, attend and spell: A neural network for large vocabulary conversational speech recognition,'' in \emph{Proc. ICASSP}, 2016, pp. 4960--4964.
\bibitem{watanabe2017hybrid} S. Watanabe, T. Hori, S. Kim, J. R. Hershey, and T. Hayashi, ``Hybrid CTC/attention architecture for end-to-end speech recognition,'' \emph{IEEE J. Sel. Topics Signal Process.}, vol. 11, no. 8, pp. 1240--1253, 2017.
\bibitem{stooke2024aligner} A. Stooke, R. Prabhavalkar, K. C. Sim, and P. M. Mengibar, ``Aligner-encoders: Self-attention transformers can be self-transducers,'' in \emph{Proc. NeurIPS}, 2024, pp. 100\,318--100\,340.
\bibitem{transformertransducer} Q. Zhang, H. Lu, H. Sak, A. Tripathi, E. McDermott, S. Koo, and S. Kumar, ``Transformer transducer: A streamable speech recognition model with transformer encoders and RNN-T loss,'' in \emph{Proc. ICASSP}, 2020, pp. 7829--7833.
\bibitem{interctc} J. Lee and S. Watanabe, ``Intermediate loss regularization for CTC-based speech recognition,'' in \emph{Proc. ICASSP}, 2021, pp. 6224--6228.
\bibitem{raffel2017monotonic} C. Raffel, M.-T. Luong, P. J. Liu, R. J. Weiss, and D. Eck, ``Online and linear-time attention by enforcing monotonic alignments,'' in \emph{Proc. ICML}, 2017, pp. 2837--2846.
\bibitem{chiu2018mocha} C.-C. Chiu and C. Raffel, ``Monotonic chunkwise attention,'' in \emph{Proc. ICLR}, 2018.
\bibitem{moritz2019triggered} N. Moritz, T. Hori, and J. L. Roux, ``Triggered attention for end-to-end speech recognition,'' in \emph{Proc. ICASSP}, 2019, pp. 5666--5670.
\bibitem{yang2022supervised} G.-P. Yang and H. Tang, ``Supervised attention in sequence-to-sequence models for speech recognition,'' in \emph{Proc. ICASSP}, 2022, pp. 7222--7226.
\bibitem{inaguma2023akd} H. Inaguma and T. Kawahara, ``Alignment knowledge distillation for online streaming attention-based speech recognition,'' \emph{IEEE/ACM Trans. Audio, Speech, Language Process.}, vol. 31, pp. 1371--1385, 2023.
\bibitem{kim2024skd} E. Kim, H. Kim, and K. Lee, ``Guiding frame-level CTC alignments using self-knowledge distillation,'' in \emph{Proc. Interspeech}, 2024, pp. 4588--4592.
\bibitem{arrnnt} J. Mahadeokar, Y. Shangguan, D. Le, G. Keren, H. Su, T. Le, C.-F. Yeh, C. Fuegen, and M. L. Seltzer, ``Alignment restricted streaming recurrent neural network transducer,'' in \emph{Proc. IEEE SLT}, 2021, pp. 52--59.
\bibitem{ctcguided} Y. Wang, Z. Chen, C. Zheng, Y. Zhang, W. Han, and P. Haghani, ``Accelerating RNN-T training and inference using CTC guidance,'' in \emph{Proc. ICASSP}, 2023, pp. 1--5.
\bibitem{shaw2018self} P. Shaw, J. Uszkoreit, and A. Vaswani, ``Self-attention with relative position representations,'' in \emph{Proc. NAACL-HLT}, 2018, pp. 464--468, doi: 10.18653/v1/N18-2074.
\bibitem{librispeech} V. Panayotov, G. Chen, D. Povey, and S. Khudanpur, ``Librispeech: An ASR corpus based on public domain audio books,'' in \emph{Proc. ICASSP}, 2015, pp. 5206--5210.
\bibitem{commonvoice} R. Ardila, M. Branson, K. Davis, M. Kohler, J. Meyer, M. Henretty, R. Morais, L. Saunders, F. Tyers, and G. Weber, ``Common voice: A massively-multilingual speech corpus,'' in \emph{Proc. LREC}, 2020, pp. 4218--4222.
\bibitem{conformer} A. Gulati, J. Qin, C.-C. Chiu, N. Parmar, Y. Zhang, J. Yu, W. Han, S. Wang, Z. Zhang, Y. Wu, and R. Pang, ``Conformer: Convolution-augmented transformer for speech recognition,'' in \emph{Proc. Interspeech}, 2020, pp. 5036--5040.
\bibitem{specaugment} D. S. Park, W. Chan, Y. Zhang, C.-C. Chiu, B. Zoph, E. D. Cubuk, and Q. V. Le, ``Specaugment: A simple data augmentation method for automatic speech recognition,'' in \emph{Proc. Interspeech}, 2019, pp. 2613--2617.
\bibitem{whisper} A. Radford, J. W. Kim, T. Xu, G. Brockman, C. Mcleavey, and I. Sutskever, ``Robust speech recognition via large-scale weak supervision,'' in \emph{Proc. ICML}, 2023, pp. 28\,492--28\,518.
\bibitem{owsm} Y. Peng, J. Tian, W. Chen, S. Arora, B. Yan, Y. Sudo, M. Shakeel, K. Choi, J. Shi, X. Chang, J.-W. Jung, and S. Watanabe, ``OWSM v3.1: Better and faster open whisper-style speech models based on e-branchformer,'' in \emph{Proc. Interspeech}, 2024, pp. 352--356.
\end{thebibliography}
\end{document}